\documentclass[epj]{svjour}
\usepackage{graphicx}

\usepackage{rotating}
\usepackage[T1]{fontenc}
\usepackage[latin1]{inputenc}

\def\lapprox{{\raise0.5ex\hbox{$<$}\hskip-0.80em\lower0.5ex\hbox{$\sim$}

}}

\def\gapprox{{\raise0.5ex\hbox{$>$}\hskip-0.80em\lower0.5ex\hbox{$\sim$}

}}

\usepackage{graphics}
\begin{document}

\title{Single-Pion Production in $pp$ Collisions at 0.95 GeV/c (I)}  

\author{
S.~Abd El-Samad\inst{8}\and%$^h$,
R.~Bilger\inst{6}\and%,%$^f$,
%A.~B\"ohm$^b$,
K.-Th.~Brinkmann\inst{2}\and%,%$^b$,
H.~Clement\inst{6} \and
M.~Dietrich\inst{6}\and%,%$^f$,
E.~Doroshkevich\inst{6}\and%,%$^f$,
S. Dshemuchadse\inst{5,2}\and%,%$^{e,b}$,
A.~Erhardt\inst{6}\and%,%$^f$,
W.~Eyrich\inst{3}\and%,%$^{c}$,
%C.~Fanara$^g$,
A.~Filippi\inst{7}\and%,%$^g$,
H.~Freiesleben\inst{2}\and%,%$^b$,
M.~Fritsch\inst{3,1}\and%,%$^{c,a}$,
R.~Geyer\inst{4}\and%,%$^d$,
A.~Gillitzer\inst{4}\and%,%$^d$,
J.~Hauffe\inst{3}\and%%$^c$,
%A.~Hassan$^h$,
%P.~Herrmann$^a$,
D.~Hesselbarth\inst{4}\and%%$^d$,
R.~Jaekel\inst{2}\and%,
B.~Jakob\inst{2}\and%,%$^b$,
L.~Karsch\inst{2}\and%,%$^b$,
K.~Kilian\inst{4}\and%,%$^d$,
H.~Koch\inst{1}\and%,%$^a$,
J. Kress\inst{6}\and%,%$^f$,
E.~Kuhlmann\inst{2}\and%,%$^b$,
S.~Marcello\inst{7}\and%,%$^g$,
S.~Marwinski\inst{4}\and%,%$^d$,
R.~Meier\inst{6}\and%,%$^f$,
%P. Michel$^e$,
K. M\"oller\inst{5}\and%,%$^e$,
H.P. Morsch\inst{4}\and%,%$^d$,
L.~Naumann\inst{5}\and%,%$^e$,
E.~Roderburg\inst{4}\and%,%$^d$,
%A. Schamlott$^e$,
P. Sch\"onmeier\inst{2,3}\and%,%$^b$,
M. Schulte-Wissermann\inst{2}\and%,%$^b$,
W.~Schroeder\inst{3}\and%,%$^c$,
M.~Steinke\inst{1}\and%,%$^a$,
F. Stinzing\inst{3}\and%,%$^c$,
G.Y. Sun\inst{2}\and%,%$^b$,
J.~W\"achter\inst{3}\and%,%$^c$,
G.J.~Wagner\inst{6}\and%,%$^f$,
M.~Wagner\inst{3}\and%,%$^c$,
U.~Weidlich\inst{6}\and%,%$^f$,
A. Wilms\inst{1}\and%,%^a$,
S.~Wirth\inst{3}\and%,%$^c$,
G.~Zhang\inst{6}\thanks{present address: Peking University}\and
%U.~Zielinski$^a$,
P. Zupranski\inst{9}%$^i$
}
%
%\offprints{H. Clement}          % Insert a name or remove this line
\mail{H. Clement \\email: clement@pit.physik.uni-tuebingen.de}
%\email{clement@pit.physik.uni-tuebingen.de}
%

\institute{
Ruhr-Universit\"at Bochum, Germany \and
Technische Universit\"at Dresden, Germany \and
Friedrich-Alexander-Universit\"at Erlangen-N\"urnberg, Germany \and
Forschungszentrum J\"ulich, Germany \and
Forschungszentrum Rossendorf, Germany \and
Physikalisches Institut der Universit\"at T\"ubingen, Auf der Morgenstelle 14,
D-72076 T\"ubingen, Germany \and
INFN Torino, Italy \and
Atomic Energy Authority NRC Cairo, Egypt \and
Soltan Institute for Nuclear Studies, Warsaw, Poland 
\\
(COSY-TOF Collaboration)}
%
%\date{\today}
%
%\date{Received: \today / Revised version: date}
\date{Received: July 10, 2006}
\abstract{
The single-pion production reactions $pp\rightarrow d\pi^+$, $pp\rightarrow
np\pi^+$  and $pp\rightarrow pp\pi^0$ were measured at a beam momentum of
0.95 GeV/c ($T_p \approx$ 400 MeV) using the short version of the COSY-TOF
spectrometer. The 
implementation of a central calorimeter provided particle identification,
energy determination and neutron detection in addition to time-of-flight and
angle measurements. Thus all pion production channels were recorded with
1-4 overconstraints. The total and differential cross sections  obtained are
compared to previous data and theoretical calculations. Main emphasis is put
on the discussion of the $pp\pi^0$ channel, where we obtain angular
distributions different from previous experimental results, however, partly in
good agreement with recent phenomenological and theoretical predictions. In
particular we observe very large anisotropies for the $\pi^0$ angular
distributions in the kinematical region of small relative proton momenta
revealing there a dominance of proton spinflip transitions associated with
$\pi^0$ $s$- and $d$-partial waves and emphasizing the important role of
$\pi^0$  $d$-waves.                 
}
\maketitle
\section{Introduction}
\label{intro}

Single-pion production in the collision between two nucleons is thought to be
the simplest inelastic process between two baryons. Therefore it came as a
surprise, when first near-threshold data on the $pp\pi^0$ channel \cite{hom}
revealed its cross section to be larger than predicted \cite{kol,mil,nis} by
nearly one order of magnitude. Meanwhile the near-threshold data base has been
much improved by exclusive measurements at TRIUMF (TINA and MINA) \cite{sta},
SATURNE (SPES0) \cite{rap} and in 
particular at the cooler storage rings CELSIUS (PROMICE/WASA)
\cite{bon,zlo,bil}, COSY (GEM, TOF) \cite{bet,DD} and IUCF \cite{iucf}, at the
latter also with polarized beam and target. However, this data base is still
far from being complete and often simple observables like angular and
invariant-mass distributions of the unpolarized cross section are still
missing. We report in this paper on measurements of such distributions at a
beam momentum of 0.95 GeV/c (corresponding to $T_p=$397 MeV) 
for the reactions $pp\rightarrow d\pi^+$, $pp\rightarrow pp\pi^0$ and
$pp\rightarrow np\pi^+$. For the latter we give here the total cross
section only, the differential distributions will be discussed in a separate
paper. 
\\

\section{Experiment}
\label{sec:2}

The measurements have been carried out at the J\"ulich Cooler Synchrotron COSY
using the time-of-flight spectrometer TOF at one of its external beam
lines. The  setup of the TOF detector system is displayed in Fig. 1. At the
entrance of the detector system the beam - collimated to a diameter smaller
than 2 mm - hits the LH$_2$ target, which has a length of 4 mm, a diameter of
6 mm and 0.9 $\mu m$ thick hostaphan foils as entrance and exit windows
\cite{has}. At a distance of 22 mm downstream of the target the two layers of
the start detector (each consisting of 0.5 mm thick scintillators cut into 12
wedge-shaped sectors) were placed followed by a two-plane fibre hodoscope
(96 x 96 fibers, 2 mm thick each ) at a distance of 165 mm from target, see
Fig. 1b. Whereas the 
start detector mainly supplies the start times for the time-of-flight (TOF)
measurements, the fibre hodoscope primarily provides a good angular resolution
for the detected particle tracks. In its central part the TOF-stop detector
system consists of the so-called Quirl, a 3-layer scintillator system  1081 mm 
downstream of the target shown in Fig. 1c and described in detail in
Ref. \cite{dah} -  and in its peripheral part of the so-called Ring, also a
3-layer scintillator system built in a design analogous to the Quirl, however,
with inner 
and outer radii of 560 and 1540 mm, respectively. Finally behind the Quirl a
calorimeter (Fig. 1a,d) was installed for identification of charged
particles and of neutrons as well as for measuring the energy of charged
particles. The calorimeter, details of which are given in Ref. \cite{kress},
consists of 84 hexagon-shaped scintillator blocks of length 450 mm, which
suffices to stop deuterons, protons and pions of energies up to 400, 300 and
160 MeV, respectively. The energy calibration of the calorimeter was performed
by the detection of cosmic muons.

\begin{figure}
\begin{center}
\includegraphics[width=21pc]{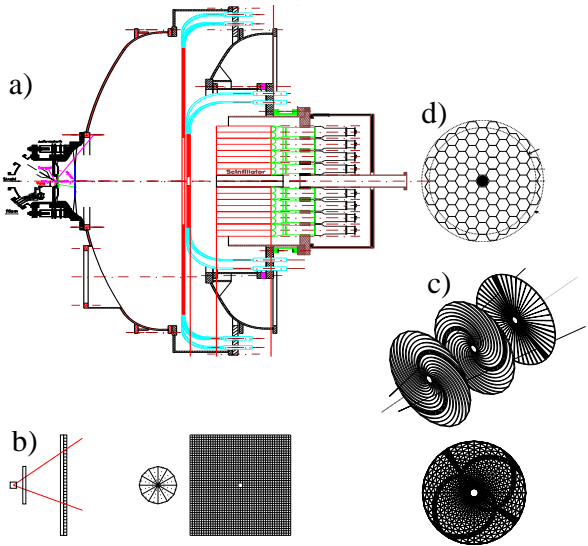}
\end{center}
\caption{Sketch of the short COSY-TOF setup used in this experiment showing
  (a) the full detector arrangement with inserts for (b) the start detector
  region with start wedges and hodoscope, (c) the central hodoscope ("Quirl")
  as the stop detector for TOF measurements and (d) the central calorimeter. } 
\end{figure}

In the experiment the trigger was set to two hits in Quirl and/or Ring
associated with two hits in the start detector.
From straight-line fits to the hit detector elements tracks of charged
particles are reconstructed. They are accepted as good tracks, if they
originate in the target and have a hit in each detector element the track
passes through. In this way the angular resolution is better than 1$^\circ$
both in azimuthal and in polar angles. If there is an isolated hit in the
calorimeter with no associated hits in the preceding detector elements, then
this hit qualifies as a neutron candidate (further criteria will be discussed
below). In this case the angular resolution
of the neutron track is given by the size of the hit calorimeter block,
i.e. by 7 - 8$^\circ$. By construction of the calorimeter a particle will hit
one or more calorimeter blocks. The number of blocks hit by a particular
particle is given by the track reconstruction. The total energy deposited by
this particle in the calorimeter is then just the (calibrated) sum
of energies deposited in all blocks belonging to the particular track.

In order to have maximum angular coverage by the detector elements and to
minimize the fraction of charged pions decaying in flight before reaching the
stop detectors, the short version of the TOF spectrometer
was used. In this way a total polar angle coverage of 3$^\circ \leq
\Theta_{Lab}\leq$ 49$^\circ$ was achieved with the central calorimeter
covering the region 3$^\circ \leq\Theta_{Lab}\leq$ 28$^\circ$. For fast
particles 
the 4\% energy resolution of the calorimeter is superior to that
from TOF measurements, the resolution of which is reduced by the short path
length. However, the TOF resolution is still much better than the $\Delta E$
resolution of the quirl elements. Hence, for particle identification, instead
of plotting $\Delta E$ versus 
$E_{cal}$, the uncorrected particle energy deposited in the calorimeter, we
utilize the relation  $\Delta E \sim (z/\beta)^2$ with the particle charge
$z=1$ and plot $1/\beta^{2}$ versus $E_{cal}$, where the
particle velocity $\beta=v/c$ is derived from the TOF measurement.

Fig. 2 shows the $1/\beta^{2}-E_{cal}$ scatterplot for two-track events. The
bands for d, p and $\pi$ are well separated. The horizontal shadow
region on the left of the deuteron band stems from deuteron breakup in the
calorimeter. Note that $E_{cal}$ in Fig. 2 is not yet corrected for 
energy and particle dependent quenching effects. By applying the
quenching correction as well as the correction for energy loss in the preceding
detector elements the kinetic energies of the detected particles are deduced.

\begin{figure}
\begin{center}
\includegraphics[width=21pc]{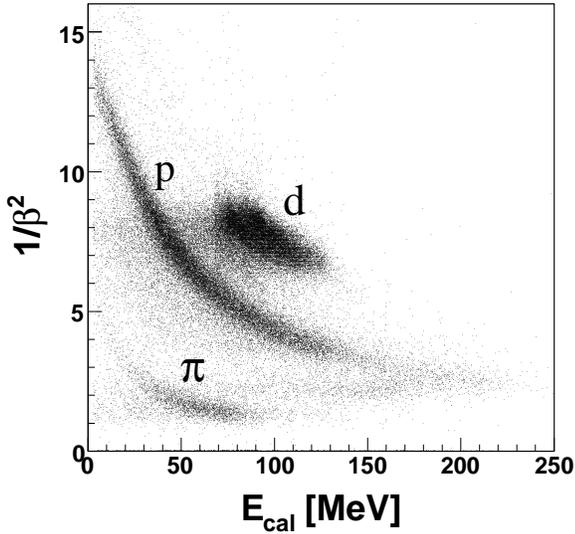}
\end{center}
\caption{$\Delta$E-E plot for particles stopped in the calorimeter. Shown is
  the first (smaller angle) track of two-track events. For the plot the
  $\Delta$E information is taken from the TOF information obtained by Quirl
  and start detectors and plotted as 1/$\beta^2$, where $\beta~=~v/c$ denotes
  the particle velocity normalized to $c$. For the E information the energy
  deposited in the calorimeter $E_{cal}$ has been taken.} 
\end{figure}

By identifying and reconstructing the two charged tracks of an event the exit
channels $d\pi^+$, $np\pi^+$ and $pp\pi^0$ can be 
separated. Kinematically the maximum  possible laboratory (lab) polar angles
are $\approx 9^\circ$ for deuterons and $\approx 32^\circ$ for protons (and
neutrons). Hence 88\% of the angular coverage for deuterons and 86\% of that
for protons from single pion production are within
the angular acceptance of the calorimeter. For charged pions the angular
coverage has been much lower with this setup, since kinematically they can
extend up to $\Theta_{Lab}~=$ 180$^\circ$. Hence within the angular coverage
of Quirl and Ring the angular acceptance for $\pi^+$ has been $\approx$40$\%$
only. Nevertheless most of the phase space part necessary for a full coverage
of the physics in single pion production has been covered (see below) by these
measurements due to the circumstance that the center-of-mass (cm) angular
distributions have to be symmetric about 90$^\circ$ because of identical
collision partners in the incident channel.

For $\Theta_{Lab}\leq$ 25.5$^\circ$ the $pp\pi^0$ events have been  
identified by requiring the two charged tracks of the events to be protons
identified in the Quirl-calorimeter system and by the condition that their
missing mass $MM_{pp}$ meets the constraints 100 MeV$/c^2$ $\leq MM_{pp}\leq$
180 MeV$/c^2$
(Fig. 3). For $\Theta_{Lab}>$ 28$^\circ$ the protons of the $pp\pi^0$ channel
have been detected in the Ring. Since kinematically they have a small $\beta$
they are easily distinguished from pions hitting the Ring and also from
elastically scattered protons. The same applies for the angular
region  25.5$^\circ > \Theta_{Lab} \leq$ 28$^\circ$, where the protons hit
only the edge of the calorimeter and in general no longer stop there. In both
cases the proton energy is calculated from the corresponding TOF measured by
the Start-Ring and Start-Quirl detector systems, respectively.
Thus the full kinematically
accessible angular range was covered for this reaction channel with exception
of the beam-hole region ($\Theta_{Lab}\leq$ 3$^\circ$).

The $d\pi^+$ events have been selected by identifying both deuteron
and pion, if both hit the calorimeter or by identifying only the deuteron, if
the pion hits the Ring. Corresponding missing mass checks have been applied. In
addition, the coplanarity condition $170^\circ < \Delta\Phi \leq 180^\circ$ is
used to further distingiush $d\pi^+$ events from three-body background. That
way even the deuteron identification in the calorimeter may be omitted, thus
allowing to check how well the deuteron breakup in the calorimeter is under
control in the Monte Carlo (MC) simulations. Within uncertainties both ways
lead to identical results.

\begin{figure}
\begin{center}
\includegraphics[width=15pc]{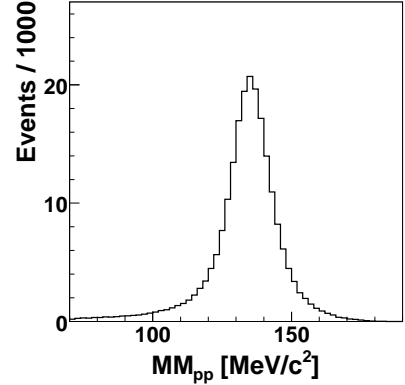}
\end{center}
\caption{Spectrum of the missing mass $MM_{pp}$ reconstructed from the
  detected proton pairs for the $pp\pi^0$ channel. } 
\end{figure}

Finally the $np\pi^+$ channel is selected by identifying proton and pion
in the calorimeter or only the proton in the calorimeter, when the second
charged track is in the Ring. In addition the missing $p\pi$ mass $MM_{p\pi}$
has to meet the condition 900 MeV$/c^2$ $\leq MM_{p\pi} \leq$ 980 MeV$/c^2$
. Also to 
suppress background from the $d\pi^+$ channel - in particular when the
deuteron is broken up and appears as a proton in the calorimeter - the $p\pi^+$
 track is required to be non-coplanar, i.e. $\Delta\Phi < 170^\circ$
 complementary to the coplanarity condition given 
above. Further on the neutron 4-momentum is reconstructed from the 4-momenta of
proton and pion and it is checked, whether a calorimeter block in the
corresponding ($\Theta$, $\Phi$) region recorded a hit accompanied without any
entries recorded in the preceding detector elements of the Quirl. If these
conditions are met, a neutron track is assumed. That way $\Theta_n$ and
$\Phi_n$ are determined by the location of this calorimeter block. Thus having
only the neutron energy undetermined experimentally we end up with 3 kinematic
overconstraints for this channel, whereas we have 1 overconstraint for the
$pp\pi^0$ channel and 4 overconstraints for the $d\pi^+$
channel. Corresponding kinematic fits were applied.

\begin{figure}
\begin{center}
\includegraphics[width=15pc]{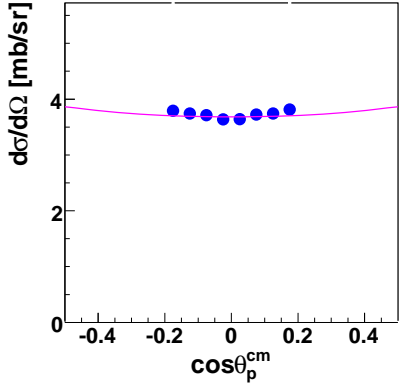}
\end{center}
\caption{Angular distribution of elastic scattering in the cm system. The data
  of elastic 
  scattering proton pairs recorded in the ring have been adjusted in absolute
  height to the SAID \cite{SAID} data base (solid line).} 
\end{figure}

The luminosity of the experiment was determined from the analysis of $pp$
elastic scattering. Due to their opening angle of $\delta_{pp}\approx84^\circ$
between both tracks, such two-track events have both hits in the Ring. They
are easily identified by using in addition the coplanarity constraint
$170^\circ < \Delta\Phi \leq 180^\circ$. Fig. 4
shows the measured angular distribution in comparison with the prediction from
the SAID database \cite{SAID}. The data have been efficiency corrected by MC
simulations of the detector setup. Adjustment of the data in absolute height
gives the required luminosity ($L = 5*10^{33}~cm^{-2}$ and $dL/dt =
3*10^{28}~cm^{-2}s^{-1}$ time averaged) and the absolute normalization of the
single-pion production cross sections, respectively, which will be discussed
in the following. 
\\

%\textbf{3. Results}
\section{Results}
\label{sec:3}

Due to the identity of the collision
partners in the entrance channel the angular distributions in the overall
center-of-mass system have to be symmetric about 90 $^\circ$, i.e. the
full information about the reaction channels is contained already in the
intervall $0^\circ\leq \Theta^{cm}\leq 90^\circ$. Deviations from this
symmetry in the data indicate systematic uncertainties in the
measurements. Hence we plot  - where appropriate - the full angular range, in
order to show the absence of major systematic errors present in our
measurement.  \\\\

%\begin{table}
%\caption{Total cross sections $\sigma_{tot}$ at $T_p~\approx~$400 MeV for the
%reactions $pp\rightarrow d\pi^+$, $pp\rightarrow np\pi^+$ and $pp\rightarrow
%pp\pi^0$ evaluated in this work and compared to previous measurements.}
%\begin{tabular}{lll}
%\hline\noalign{\smallskip}
%$pp\rightarrow d\pi^+$ & $pp\rightarrow np\pi^+$ &  $pp\rightarrow pp\pi^0$ \\
%0.74 (2) & 0.47 (2) & 0.100 (7) \\
%0.72$^{a)}$ &  & 0.092 (7)$^{b)}$ \\
%  &  & 0.070 (7)$^{c)}$\\
%\end{tabular}

Tab. 1: Total cross sections $\sigma_{tot}$ at $T_p~\approx~$400 MeV for the
reactions $pp\rightarrow d\pi^+$, $pp\rightarrow np\pi^+$ and $pp\rightarrow
pp\pi^0$ evaluated in this work and compared to previous measurements.

\vspace{0.5cm}
%\begin{tabular}{l|lll} 
\begin{tabular}{llll} 
\hline

 & ~~~~~~~~~~~~~~~~~~& $\sigma_{tot}$ [mb]~~~~~~~~~~\\

 & $pp\rightarrow d\pi^+$ & $pp\rightarrow np\pi^+$ & $pp\rightarrow pp\pi^0$\\

\hline

& 0.74 (2) & 0.47 (2) & 0.100 (7)\\

 & 0.72$^{a)}$ &  & 0.092 (7)$^{b)}$\\ 

 &  &  & 0.070 (7)$^{c)}$\\

\hline

 \end{tabular}\\

 $^{a)}$ Ref. \cite{SAID} ~~~  $^{b)}$ Ref. \cite{bil} ~~~ $^{c)}$
 Ref. \cite{sta}\\

\vspace{1cm}

%\vspace{1cm}

The evaluated total cross sections for the three channels are given in Tab. 1
together with previous results. The uncertainties assigned 
are based on systematics for acceptance and efficiency corrections as obtained
by variation of MC simulations for the detector response, where we have varied
the MC input assuming either pure phase space or some reasonable models for
the reaction under consideration. Statistical
uncertainties are negligible compared to the systematic uncertainties.\\

\subsection{$pp\rightarrow d\pi^+$}
\label{sec:3.1}

Absolute and differential cross sections for this reaction channel are very
well known from previous experiments. Hence we use the analysis of our data
for this channel primarily as a check of the reliability of our measurement
and data analysis. In Fig. 5 our results for the $\pi^+$ angular distribution
are shown in comparison with the prediction from the SAID data base
\cite{SAID}. Since in the measurement we cover only angles
$\Theta_d^{cm}<$90$^\circ$ we show in Fig.5 only the appropriate half of the
full angular distribution. Note, however, that due to the symmetry
of the angular distribution around $cos\Theta_d^{cm}=0$ most of the physically
relevant phase space part has been covered in this measurement. 
We find good agreement with the SAID data base both in absolute magnitude
and in the shape of the angular distribution. Since the
pions of this channel cover the angular range of both Quirl and Ring, i.e.,
stem from Quirl-Quirl and Quirl-Ring coincidences, the good agreement with
SAID assures that there are no significant problems with correlating the
efficiences of Quirl and Ring.
\\

\begin{figure}
\begin{center}
\includegraphics[width=15pc]{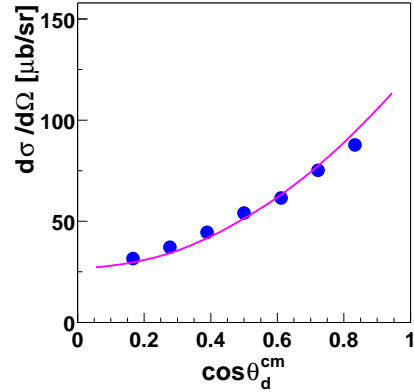}
\end{center}
\caption{Pion angular distribution in the cm system for the reaction
  $pp\rightarrow 
  d\pi^+$. The data of this work ( full circles) are compared to 
% previous results \cite{aeb} ( solid boxes and crosses) and 
  the SAID \cite{SAID} data base (solid line).}
\end{figure}

\subsection{$pp\rightarrow pp\pi^0$}
\label{sec:3.2}

This channel has received increasing attention since first
measurements in the threshold region \cite{iucf} at IUCF and later also at
CELSIUS \cite{bon} uncovered the total
cross section to be nearly an order of magnitude larger than predicted
theoretically \cite{kol,mil,nis}. Very recent close-to-threshold measurements
at COSY-TOF revealed the experimental total cross sections to be even larger
by roughly 50$\%$  \cite{DD} than previously measured. There it was shown that
the pp final state interaction (FSI) has a very strong influence on the
reaction dynamics close to threshold with the
consequence that a substantial part of the cross section is at small lab
angles, which were missed in IUCF and CELSIUS measurements near
threshold. At higher energies, 320 MeV $\leq T_p \leq$ 400 MeV, where the
influence of the pp FSI decreases more and more, the
total cross section data measured at TRIUMF \cite{sta}, SATURNE \cite{rap},
COSY-GEM \cite{bet} and CELSIUS (PROMICE/WASA) \cite{bil} are in agreement
with each other with the exception of 
a 20$\%$ discrepancy at $T_p \approx$ 400 MeV between CELSIUS and TRIUMF
results. Our value, see Tab. 1, is in agreement with the CELSIUS result.
Recent measurements with polarized beam \cite{rap} and partly also polarized
target \cite{hom} added much to the detailed knowledge of this reaction from
threshold up to $T_p\approx$ 400 MeV. 

\begin{figure}
\begin{center}
\includegraphics[width=20pc]{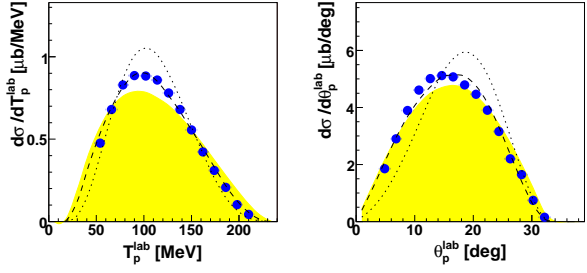}
\end{center}
\caption{Spectra for kinetic energies and polar angles (lab system) of
  detected proton 
  pairs stemming from the  $pp\rightarrow pp\pi^0$ reaction. The shaded areas
  show the corresponding phase space distributions for comparison. Dashed and
  dotted lines give calculations with the ansatz eq. (4) and the ansatz of
  Ref. \cite{zlo}, i.e. eq. (3), respectively. }
\end{figure}

Despite the wealth of experimental information on this reaction there remain 
a number of problems, which are not yet sufficiently settled. E.g., the
anisotropy of the pion angular distribution in the overall cm system,
characterized 
by the anisotropy parameter $b$, which traditionally \cite{sta,rap} is defined
by 
\begin{center}
$\sigma (\Theta_{\pi^0}^{cm})~\sim ~ 1/3 + b*
cos^2\Theta_{\pi^0}^{cm},~~~~~~~~~~~~~~~~(1)$ \\ 
\end{center}
where $\Theta_{\pi^0}^{cm}$ denotes the $\pi^0$ polar angle in the overall
cm system, shows a big scatter in the results from different measurements, see,
e.g., Fig.11 in Ref.\cite{DD} for $T_p \leq$ 400 MeV and Fig.6 in
Ref.\cite{rap} for higher incident energies. Note that in Ref.\cite{DD} the
Legendre coefficient  $a_2$ for p-waves is plotted according to the ansatz
\begin{center}
$\sigma (\Theta_{\pi^0}^{cm})~\sim ~ 1 + a_2 * 
(3~cos^2\Theta_{\pi^0}^{cm} - 1) / 2,~~~~~~(2)$ \\ 
\end{center}

In this paper we will use eq. (2) analogously also for fitting the
experimental proton angular distributions $\sigma(\Theta_p^{pp)}$ by replacing
$\Theta_{\pi^0}^{cm}$ 
with $\Theta_p^{pp}$, the proton angle in the pp subsystem.
The parameters $b$ and $a_2$ are related by 
$a_2 = 2b/(1+b) \approx 2b$ for $b << 1$. We will use the
quantity $a_2$ in the following discussion of the angular distributions. 

Near-isotropy is found for  $T_p <$ 400 MeV with $a_2$ staggering between -0.1
and +0.1 - with a tendency for negative values. The latter would mean
that d-wave contributions inducing negative $a_2$ 
values are already present close to threshold \cite{zlo}. A clearer trend
towards positive $a_2$ values is observed for  $T_p >$ 400 MeV.

As we will show below, our results 
for $a_2$ are at variance with previous results and hence need a more detailed
consideration. To this end we start the presentation and discussion of our
results first with energy and angular distribution of the protons
in the lab system as displayed in Fig.6.  In this and in the following
one-dimensional figures
phase space distributions are shown by shaded areas for comparison. 
%We see
%that different from previous investigations most of the reaction phase space
%has been covered in our measurement. 
We see that the data do not deviate
vigorously from the phase space distributions, as we would expect, e.g., if
$\Delta$ excitation would play a dominant role in this reaction channel. This
is also visible in the 
experimental Dalitz  plots of $M_{p\pi^0}^2$ versus $M_{pp}^2$ and
$M_{p\pi^0}^2$ versus $M_{p\pi^0}^2$ displayed in Fig.7. The data cover 
%- with the exception of the beam-hole region - 
essentially the full available elliptic phase
space areas and yield distributions, which are close to flat 
with just one pronounced excursion in the region of the $pp$ FSI. This is
reflected also in the projections of the Dalitz plot leading to the spectra of 
the invariant masses $M_{p\pi^0}$ and $M_{pp}$ (Fig.8). The latter exhibits a
small spike at the $pp$ threshold due to $pp$ FSI, though in total its
influence is of minor importance as expected from the small amount of s-wave
between the two protons being available at this energy \cite{iucf}. In the
$M_{p\pi^0}$ spectrum we compare our data with the ones from CELSIUS as given
in Ref.\cite{bil}. We find agreement between both data sets with the exception
of the region around $M_{p\pi^0} \approx$ 1100 MeV/$c^2$, where we obtain a
%20$\%$ 
somewhat larger yield.

\begin{figure}
\begin{center}
\includegraphics[width=20pc]{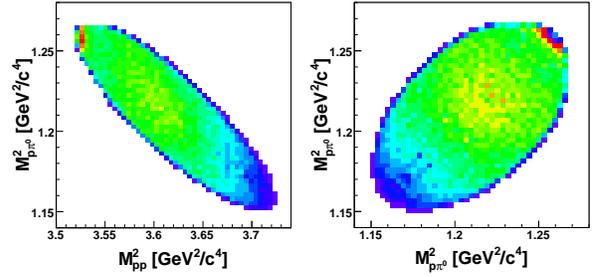}
\end{center}
\caption{Dalitz plots for the invariant mass combinations  $M_{p\pi^0}^2$ 
  versus $M_{pp}^2$ and $M_{p\pi^0}^2$ versus $M_{p\pi^0}^2$ as obtained
  from the data for the $pp\rightarrow pp\pi^0$ reaction. Note that the plots
  are efficiency but not acceptance corrected, hence the tiny deviations from
  the elliptic circumference at the upper corners due to the excluded beam-hole
  region.}
\end{figure}

%\includegraphics[width=30pc]{f9_picm_ppsubsyst_wasa.eps}
%\end{center}
%\caption{Angular distributions of pions (overall cms) and protons ( pp
%  subsystem, Jackson frame) for the $pp\rightarrow pp\pi^0$

\begin{figure}
\begin{center}
\includegraphics[width=20pc]{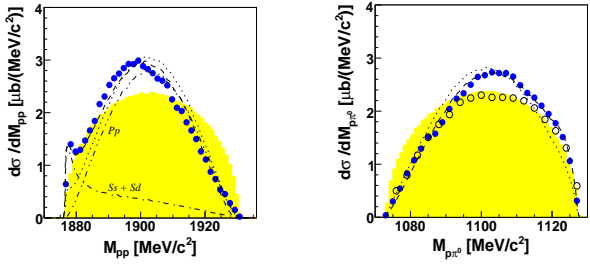}
\end{center}
\caption{Differential cross sections in dependence of invariant masses
  $M_{pp}$ and $M_{p\pi^0}$ for the $pp\rightarrow pp\pi^0$
  reaction. Data of this work are shown by full circles, the ones from
  Ref. \cite{bil} by open circles and phase space by the shaded area. For the
  explanation of dashed and dotted lines see caption of Fig. 6. The
  dash-dotted lines in the $M_{pp}$ distribution show the partial
  wave contributions $Ss + Sd$ and $Pp$ as obtained from eq. (4).} 
\end{figure}

Now we turn again to the angular distributions, which are shown in Fig.9, on
the left for the pions denoted by their cm polar angles
$\Theta_{\pi^0}^{cm}$ and on the right for protons in the $pp$ subsystem
(Jackson frame) denoted by $\Theta_{p}^{pp}$, i.e. we use the
same coordinate system scheme as defined in the IUCF publication
\cite{iucf}. Both 
distributions are close to flat, exhibit, however, a clearly negative
anisotropy parameter with $a_2=-0.12(1)$ for pions and $a_2=-0.10(1)$ for 
protons. 

%\textbf{4. Discussion of Results}
\section{Discussion of Results}
\label{sec:4}

The negative anisotropy parameter observed in this experiment for the $\pi^0$
angular distribution comes as a surprise, since all previous
experiments around $T_p\approx$ 400 MeV gave - or indicated at least - a
positive value for the pions, the most serious discrepancy being with the
PROMICE/WASA results \cite{bil} of $a_2=+0.127(7)$ (note that the $b$ values
given in \cite{bil} need to be divided by a factor of 3, in order to comply
with our definition of $b$ in eq. 1), since this measurement provided the best
statistics and phase space coverage of all previous experiments at this
energy. Although this work reports negative $a_2$ 
values for $T_p~\leq$ 360 MeV, positive values are found for
$T_p~\geq$ 400 MeV. 

\begin{figure}
\begin{center}
\includegraphics[width=20pc]{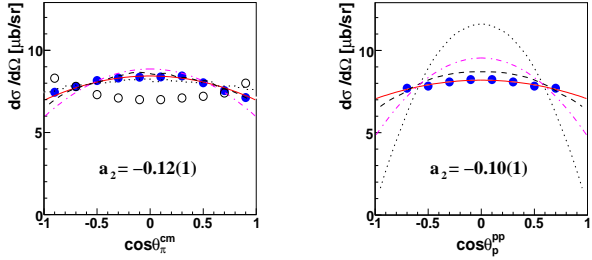}
\end{center}
\caption{Angular distributions of pions (overall cm system) and protons ( pp
  subsystem, Jackson frame) for the $pp\rightarrow pp\pi^0$
  reaction. Data of this work are shown by full circles, the fit to the data
  with eq. (2) by solid lines, the results of Ref. \cite{bil} by open circles
  and the prediction of Ref. \cite{han2} by the
  dash-dotted curve. For the
  explanation of dashed and dotted lines see caption of Fig. 6.}
\end{figure}

From their analysis of polarization data the authors of Ref. \cite{iucf} also
find some predictions for the unpolarized angular distributions, though with
very large uncertainties. For protons they get $a_2 = -0.34(81)$ and for pions
$a_2 = 0.17(11)$. The first one agrees in sign and value with our
results, however, their value for the pions has an opposite sign.
Nevertheless, since these numbers were obtained only indirectly with very
large uncertainties, this is not a point of major concern -
in particular since $\pi^0$ $d$-waves , which as we demonstrate in this paper
are vital for a proper understanding of the reaction, are not taken into
account in Ref. \cite{iucf}. We
also note that a recent measurement of this reaction at $T_p$ = 400 MeV was
carried out at CELSIUS-WASA, too. The results of its analysis also
provide a negative $a_2$ parameter for the pion angular distribution
\cite{kel,pia}. If we restrict our data to the same proton angular range
covered in the CELSIUS-WASA experiment, then we find full consistency between
both results. 

On the theoretical side extensive meson-exchange calculations as well as
partial wave analyses were carried out very recently by the J\"ulich
theory group \cite{han1,dhh,han2}. Their calculations
partly are in good agreement with the polarization data. Interestingly their
prediction for proton and pion angular distributions (shown in Fig. 9 by
the dash-dotted lines) also results in negative $a_2$ agreement with our
results.  
%They also find that the major difference between their
%calculations and their phase shift analysis, which essentially resides on the
%IUCF data. 
More advanced calculations based
on chiral pertubation theory are in progress. First steps in this direction
have already been taken by this group \cite{len}.

\begin{figure}
\begin{center}
\includegraphics[width=15pc]{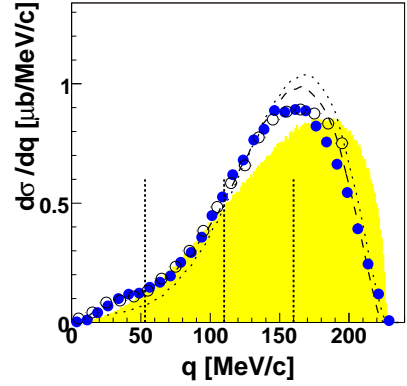}
\end{center}
\caption{Differential cross section in dependence of $q$, which is half of the
  relative momentum between the proton pair ($q^2 = M_{pp}^2 /4 - m_p^2$). 
  Data of this work (full circles) are compared to
  results from Ref. \cite{bil} (open circles) and to phase space (shaded
  area). For ease of comparison the data from Ref. \cite{bil} have been
  normalized to the maximum of our distribution. For the
  explanation of dashed and dotted lines see caption of Fig. 6.}
\end{figure}

In order to get a better insight into this problem and to compare in more
detail with the CELSIUS results we next consider the angular
distributions in dependence of $q$, defined as half of the relative momentum
between the two protons and given by $q^2 = M_{pp}^2 /4 - m_p^2$. In this
definition $q$ then denotes the momentum of a proton in the $pp$
subsystem. The 
distribution obtained from our data is compared in Fig.10 with the one given in
Ref. \cite{bil}. Whereas good agreement between both results is found for
$q~\leq~150$ MeV/c, substantial deviations appear for larger $q$ values. In
general large $q$ values are associated with large opening angles between the
protons, which in our experiment are fully covered. However,
with the setup at PROMICE/WASA, where protons were detected only for lab
angles smaller than 20$^\circ$, the high-$q$ region was not covered as
completely as at TOF and substantial extrapolations had to be applied in the
acceptance correction for this part of the CELSIUS data.\\

\begin{figure}
\begin{center}
\includegraphics[width=20pc]{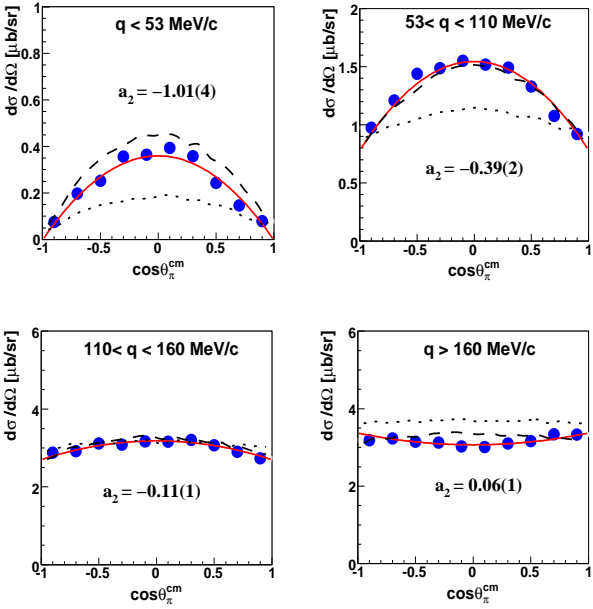}
\end{center}
\caption{Pion angular distributions (overall cm system) for 4 different
  $q$-ranges as 
  indicated for the $pp\rightarrow pp\pi^0$
  reaction. Data from this work are shown by full circles, the fit to the data
  with eq. 2 by solid lines. For the
  explanation of dashed and dotted lines see caption of Fig. 6.  }
\end{figure}

%\textbf{(i) $q$-dependence of angular distributions}
\subsection{$q$-dependence of angular distributions}
\label{sec:4.1}

In Figs. 11 and 12 we plot pion and proton angular distributions for the four
different $q$ regions indicated in Fig. 10. The corresponding anisotropy
parameters $a_2$ obtained by fitting eq. (2) to the data are indicated in each
of the plots. We see that the pion angular
distribution is strongly anisotropic with  $a_2 = -1$ for the lowest $q$
range, where there are also results from Ref. \cite{bil} - and in
fact they also exhibit a very strong anisotropy with  $a_2~<~0$. With
increasing $q$ the pion angular distributions are getting gradually flatter
and for large $q$ the $\Theta_{\pi^0}^{cm}$ distribution gets even
curved with $a_2~>~ 0$.  

The lowest $q$ range ($q < $ 53 MeV/c) is particularily interesting. It
corresponds to a kinetic energy of the protons in the $pp$ subsystem of
$T_p^{pp}$ = $q^2 / m_p < $ 3 MeV, a constraint, which has been frequently
used to have the $pp$ subsystem safely selected in a relative $S$-wave
state. That way especially simple configurations in the exit channel are
selected  allowing a deeper insight into basic reaction mechanisms of $\pi^0$
production. In particular, configurations are selected that way, where $pp
\rightarrow \Delta N(l=1) \rightarrow pp\pi^0$, i.e.,  which have the
$\Delta N$ system in relative p-wave in the intermediate state. 
In fact, the observed exceptionally large pion anisotropy of $a_2 = -1$
equivalent to a  pure $sin^2\Theta_{\pi^0}^{cm}$ distribution means that at
$T_p$ = 400 MeV the 
$\pi^0$ production process associated with $S$-wave protons in the final
state happens to be a pure proton spinflip process originating from the
transitions $^3P_0 \rightarrow ^1S_0s$ and  $^3P_2 \rightarrow ^1S_0d$, where
capital letters refer to partial waves 
in the pp system and small letters to partial waves of $\pi^0$ relative
to the pp system (see 
e.g. Table I in Ref. \cite{iucf} for a list of contributing partial
waves). In fact, this very special situation has been predicted in the
phenomenological model of Ref. \cite{zlo}, which - based on the CELSIUS-WASA
measurement at $T_p$ = 310 MeV - gave already a good description of the
RCNP data \cite{rcnp} at $T_p$ = 300, 318.5, 345 and 390 MeV in the $T_p^{pp} <
$ 3 MeV range. If we use this model ansatz the cross section is given by 

\begin{center}
$\sigma / PS ~\sim ~ FSI  [ A_0^2 + ( 2A_0B_0 \tilde k^2 + B_0^2 \tilde k^4 ) 
 cos^2\Theta_\pi^{cm} ]$\\
 $~~~~~~~~~~~~~~~~~+ ~\tilde q^2  [ C^2 + D^2 \tilde k^2 sin^2\Theta_{p}^{pp} ], ~~~~~~~~~~~~~~~~(3)$
\end{center}

where PS and FSI stand for phase space and FSI factors ( for the latter see
eq. (2) of Ref. \cite{zlo}). The pion cm momentum $\tilde k$ and the momentum
$\tilde q$, which is half the relative momentum between the
proton pair, are given here in units of the pion mass. $A_0$ and $B_0$ denote
pion $s$- and $d$-wave 
amplitudes for the transitions $^3P_0\rightarrow ^1S_0s$ and  $^3P_2
\rightarrow ^1S_0d$ - as defined in Ref. \cite{zlo} with $B_0$ = -1.2 $A_0$ -
whereas $C^2$ = 0 and $D$ = 0.98 $A_0$ stand for proton $P$-wave
contributions of the transitions $^1S_0\rightarrow ^3P_0s$, $^3P_0\rightarrow
^3P_1p$ and $^3P_1\rightarrow ^3P_jp$ with $j=0,1,2$, respectively. 

If we use eq. (3) together with the parameters $A_0,~B_0$, $C^2$ and $D^2$ as
determined in Ref. \cite{zlo}, then we 
obtain the dotted curves in Figs. 6 and 8 - 12. The agreement of these
calculations with our data is striking for the lowest $q$ bin (Figs. 11, 12) 
if renormalized to the same absolute cross section of this bin. Also for the
other differential distributions it is partly surprisingly good -
with the exception of the proton angular distributions, where the
$sin^2\Theta_{p}^{pp}$ ansatz badly 
fails. However, the trend in the pion angular distribution is correctly
reproduced. We return to this point at the end of this chapter, section (ii). 

For the pion angular distribution in the lowest-$q$ region a very strong
anisotropy with  $a_2~<~ 0$ has recently also been observed in COSY-ANKE
measurements at $T_p$ = 800 MeV \cite{dym}. In a  
very recent work by Niskanen \cite{nis1} these anisotropies are explained by a
strong energy dependence of the forward cross section due to interfering pion
partial waves.

For the $\Theta_{p}^{pp}$ proton angular distribution,
unfortunately, no data are shown in Ref. \cite{bil} to compare with (there the
proton angular distribution in the overall cm system is shown instead, which is
slightly different from the one in the $pp$ subsystem). In the low-$q$ region,
i.e. in the region affected most strongly by the $pp$ FSI, we expect the
$^1S_0$ partial wave between the two protons to dominate. Indeed, we find the
$\Theta_{p}^{pp}$ angular distribution to be compatible with isotropy within
uncertainties. With increasing $q$ the distribution gets more and more
anisotropic with  $a_2~<~ 0$.

\begin{figure}
\begin{center}
\includegraphics[width=20pc]{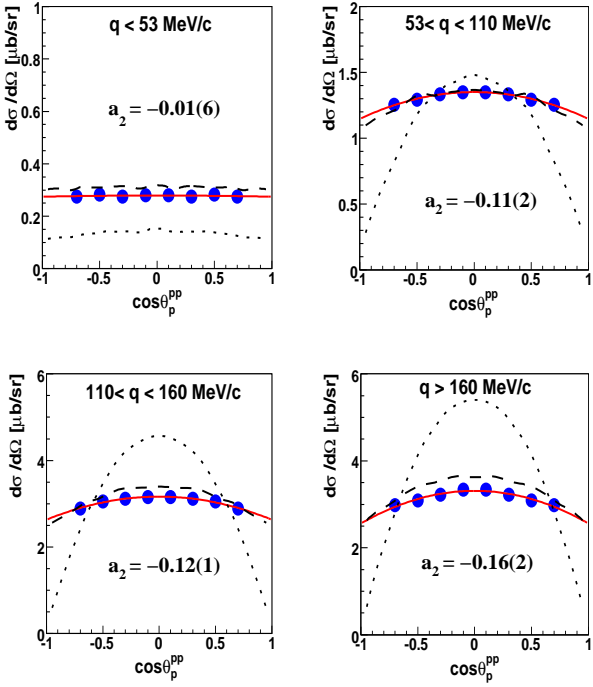}
\end{center}
\caption{Proton angular distributions ($pp$-subsystem, Jackson frame) for 4
  different $q$-ranges as indicated for the $pp\rightarrow pp\pi^0$
  reaction. Data of this work are shown by full circles, the fit to the data
  with eq. 2 by solid lines. For the
  explanation of dashed and dotted lines see caption of Fig. 6. }
\end{figure}

The observed trend in the angular distributions is not
unexpected. In the low-$q$ region the $pp$ system has kinematically the least
internal freedom and hence no chance to develop much dynamics involving
higher partial waves. At the same time the pion has kinematically the largest
freedom within the $pp\pi^0$ system with the possibility to involve
dynamically higher partial waves, which then show up in appreciable
anisotropies of the pion angular distributions. At large $q$ the situation is
reversed and the pions are kinematically bound to low partial waves, i.e. flat
angular distributions. The observed anisotropies with negative $a_2$ values 
both for proton and pion angular distributions in the lower $q$ range point to
the importance of proton spinflip transitions in this process, which are
associated with $\Delta$ excitation in the intermediate state as well as with
$\pi^0$ $s$- and $d$-waves in the exit channel. Our results are in support
of the conclusions in Ref. \cite{zlo} that $\pi^0$ $d$-waves play an important
role already close to threshold, obviously favored by the possibilty of
exciting the $\Delta$ in this manner.

%\textbf{(ii) partial-wave description of data}
\subsection{partial-wave description of data}
\label{sec:4.2}

Finally we come back to the partial-wave ansatz in Ref. \cite{zlo}, which is a
simplified version of the more comprehensive ansatz in Ref. \cite{bil} and
which we have seen to work quite well for the pion angular distributions,
however, failing badly for the proton distributions. In order to overcome
these shortcomings we modify eq. (3) slightly (and more plausible, if we look
at the partial-wave expansions given in Ref. \cite{bil,iucf}) by

\begin{center}
$\sigma / PS ~\sim ~ FSI  [ A_0^2 + ( 2A_0B_0 \tilde k^2 + B_0^2 \tilde k^4 ) 
 cos^2\Theta_\pi^{cm} ]$\\
 $~~~~~~~~~~~~~~~~~~~~~~~~~~~+ ~\tilde q^2  [ C^2 + \tilde k^2 D_1^2 ( 1 + D_2 ( cos^2\Theta_{p}^{pp} - 1/3
 ))], ~~(4)$ 
\end{center}

where the latter coefficients for the $Pp$ waves are related to the one in
eq. (3) and in Ref. \cite{zlo} by $D^2$ = $2/3 D_1^2$ in combination with
$D_2=-3/2$. Note that in this ansatz $Ps$ and $Pp$ contributions are taken
into account only very rudimentarily in the hope that still their most
important parts are covered by the ansatz. Also for the $Sd$ contribution a
$cos^4\Theta_\pi^{cm}$ term has been neglected due to its smallness and in
order to keep the ansatz as close as possible to the one of Ref. \cite{zlo}.
%With regard to the nomenclature used in Ref. \cite{bil} we have
%the following relations: 
%\begin{center}
%$C_{Ss}^2 = A_0^2$, $C_{Sd}^2 = B_0^2$ with $C_{Sd}^{k2}=3$,
%$C_{Ps}^2 = C$, $C_{Pp}^2 = D_1$ and $C_{Pp}^{q2}=D_2$.
%\end{center}

Since the model ansatz of Ref. \cite{zlo} works already very well for the pion
angular distributions, we do not touch the pion $s$- and $d$-wave parts,
i.e. leave the correlation  $B_0$ = -1.2 $A_0$, take $A_0$ as a general scale
parameter to reproduce the integral cross section and adjust $C^2,~D_1^2,~D_2$
for best reproduction of the proton angular and $q$ distributions. As a result
we obtain 
\begin{center}
 $B_0$ = -1.2 $A_0$, $C^2$ = 0, $D_1^2$ = 0.22 $A_0^2$ and
 $D_2$ = -0.34 .
\end{center}
Actually we get a slightly better description having $C^2$ =
-0.006. However, since $C^2$ as a squared quantity should not be negative, we
set $C^2$ = 0 as was done also in Ref. \cite{zlo}.
The resulting values mean that aside from $Ss$ and $Sd$ contributions the
dominant contribution 
comes from $Pp$ configurations, whereas $Ps$ configurations are of minor
importance. The decomposition of the total cross section into interfering $Ss$
and $Sd$ wave contributions on the one hand and $Pp$ contributions on the other
hand is shown in Fig. 8 for the $M_{pp}$ invariant mass distribution. We
see that the $Ss + Sd$ wave part adds significantly to
the total cross section and accounts very well for the observed FSI effect in
the $M_{pp}$ invariant mass distribution.
Note that $Sd$ contributions, which turn out here to be crucial for
the understanding of the neagative cuirvature of the pion angular
distributions, have not been taken into account in the analysis of
Ref. \cite{iucf}.  

Though this ansatz is still very simple compared to the full
partial-wave ansatz as given in Refs. \cite{bil,iucf}, it is obviously
sufficient to provide a near quantitative 
description of the data both for angular distributions and invariant
masses (dashed lines in Figs. 6, 8 -12). It is not the
aim of this work to provide a perfect partial-wave fit to the data. We rather
put here the main emphasis on revealing the dominating partial waves in this
reaction, which are responsible for the main signatures in the (unpolarized)
differential observables. Compared to  Ref. \cite{zlo} the only
major change is the replacement of the $D^2$ term with its inappropriate
$sin^2\Theta_{p}^{pp}$ dependence  by the $D_1^2$ and $D_2$ terms, which
provide a more appropriate angular dependence for the protons. This
modification, however, is not 
surprising, since the proton angular dependence has actually not been tested 
in  Ref. \cite{zlo}. From the successful description of our data as well of
those treated in Ref. \cite{zlo} we conclude that the ansatz eq. (4) provides
an amazingly successful description of the ( unpolarized) $pp \rightarrow
pp\pi^0$ data from threshold up to $T_p$ = 400 MeV.
\\

%\textbf{5. Summary}
\section{Summary}
\label{sec:5}

It has been demonstrated that by addition of the central calorimeter the
COSY-TOF setup is capable of providing a reliable particle identification on
the basis of the $\Delta$E-E technique. In this way the different single pion
production channels were separated. The results for the $d\pi^+$ 
channel agree well with previous results. For the $pp\pi^0$ channel significant
deviations from previous investigations were obtained for angular
distributions as well as for invariant mass spectra. It has been demonstrated,
that pions and protons exhibit $q$-dependent angular distributions. Hence a
full coverage of the phase space appears to be mandatory for reliable
experimental results on this issue. For the
lower $q$ region angular distributions with negative $a_2$ parameter dominate
pointing to the importance of proton spinflip transitions associated with
$\pi^0$ $s$- and $d$-waves. In
particular,
 we observe a pure $sin^2\Theta$ distribution for $T_p^{pp} <$ 3 MeV,
which derives from a special combination of the spinflip transitions  $^3P_0
\rightarrow ^1S_0s$ and  $^3P_2 \rightarrow ^1S_0d$. Different
from previous experiments this has been the first measurement at $T_p~\approx$
 400 MeV covering practically the full reaction phase space. The data thus may
 serve as a reliable basis for a comprehensive phase shift analysis of this
 reaction. 
\\

\begin{acknowledgement}

This work has been supported by BMBF, DFG (Europ. Gra\-duiertenkolleg
683) and COSY-FFE. We acknowledge valuable discussions with Murat Kaskulov, Pia
Thorngren-Engblom, Jo\-zef Z{\l}omanczuk, Christoph Hanhart and Alexander
Sibirtsev. 
\end{acknowledgement}

\end{document}